\documentclass[sigconf,nonacm]{acmart}

\setcopyright{none}
\renewcommand\footnotetextcopyrightpermission[1]{}

\usepackage{amsmath}
\usepackage{booktabs}
\usepackage{tabularx}
\usepackage{graphicx}

\AtBeginDocument{%
  }

\begin{document}

\title{A User-Centric Context-Aware Permission Governance Framework for Privacy Control in Default Mobile Applications}

\author{Asmau Yetunde Adeniran}
\authornote{Corresponding author.}
\email{asmauadeniran2004@gmail.com}
\affiliation{%
  \department{Department of Cybersecurity}
  \institution{Air Force Institute of Technology}
  \city{Kaduna}
  \country{Nigeria}}

\author{Adeniran Kolade Ademuwagun}
\affiliation{%
  \department{Department of Cybersecurity}
  \institution{Air Force Institute of Technology}
  \city{Kaduna}
  \country{Nigeria}}

\author{Fatimah Adamu-Fika}
\affiliation{%
  \department{Department of Cybersecurity}
  \institution{Air Force Institute of Technology}
  \city{Kaduna}
  \country{Nigeria}}

\author{Samaila Musa Abdullahi}
\affiliation{%
  \department{Department of Cybersecurity}
  \institution{Air Force Institute of Technology}
  \city{Kaduna}
  \country{Nigeria}}

\author{Freeman Bitrus}
\affiliation{%
  \department{Department of Cybersecurity}
  \institution{Air Force Institute of Technology}
  \city{Kaduna}
  \country{Nigeria}}

\author{Fortune Daberechi Ifeanyi}
\affiliation{%
  \department{Department of Cybersecurity}
  \institution{Air Force Institute of Technology}
  \city{Kaduna}
  \country{Nigeria}}

\renewcommand{\shortauthors}{Anonymous}

\begin{abstract}
Mobile operating systems offer runtime permission
systems that are meant to improve the user control over access to
sensitive data. Nevertheless, default or pre-built mobile applications
are commonly much more deeply embedded into the system and can run with
high privilege and cannot be scrutinized by the user. Current models of
permission normally permit access in a persistent or a temporary state
as long as an application session is alive, but do not differentiate
between different features of the application. Consequently, users can
give authorizations with little understanding of at what time and why
certain information is being accessed. This paper presents a
context-sensitive and user-focused permission governance system that
would enhance privacy management in default mobile applications. The
framework proposes a feature-based authorization option known as allow
when needed which only allows access to a data to the functionality that
needs it and not to the entire application session. A weighted scoring
system was created to determine the privacy implication of the choices
made by users, which was founded on the sensitivity of permission and
the type of authorization. To simulate 30 realistic permission request
situations in six default application types that are commonly used, a
web-based simulation platform used to enable controlled, early-stage
evaluation prior to native implementation platform was put in place. The
assessment methodology adopts an exploratory approach using a
cross-sectional survey comprising of 104 respondents to test the privacy
awareness and behavior concerning the default applications and to
supplement with formative usability testing consisting of eight
respondents to test the interaction with the prototype system. According
to the results of the surveys, not all users always elaborate the
default app permission and would prefer that it be explained in a more
contextual way before they could adopt it. These findings provide
preliminary evidence that context-aware permission governance can
improve user understanding and decision clarity during permission
interactions. Given the limited availability of user-centric evaluations
specifically targeting default mobile applications, this study adopts an
exploratory simulation-based approach as an appropriate first step
toward examining feature-level authorization and privacy feedback
mechanisms before native mobile deployment.

\end{abstract}



\maketitle
\pagestyle{plain}

\section{Introduction}

Mobile phones are now vital computing gadgets of communication,
navigation, access to information, health care, and daily digital
transactions. Most of this is through default mobile applications
included with the operating system, including browsers, messaging apps,
maps, cameras, phone services, and health utilities. The default
applications often seek permission to access sensitive resources, such
as location, microphone, camera, contacts, storage and notifications in
order to deliver these services. Although this kind of access
facilitates convenience and usability, it also poses privacy threat
where the access may not be easily visible, restricted, and interpreted
by the user. Felt et al.~examined user attention to and comprehension of
mobile platforms and
reported that users often struggle to accurately evaluate the privacy
implications associated with application permissions~\cite{felt2012android}.

Mobile operating systems attempt to mitigate these risks through
permission-based access control mechanisms, which allow users to grant
or deny applications access to protected system resources. Over time,
these systems have evolved from install-time permission models toward
runtime permission systems that allow users to make authorization
decisions during application use. Wijesekera et al.~demonstrated that
permission decisions depend on the context in which access occurs,
emphasizing the importance of contextual and understandable permission
prompts~\cite{wijesekera2015remystified}.

Despite these improvements, existing permission systems largely operate
at the application or session level rather than the feature level. Once
a permission is granted, access may persist for the duration of the
application session even when the specific feature that required the
permission is no longer active. Mendes et al.~conducted a field study
examining user expectations regarding mobile application privacy and
found that many permission accesses occur in ways that users do not
anticipate, resulting in discrepancies between user expectations and
system behavior~\cite{mendes2022expectation}.

Privacy concerns are particularly significant for default or
pre-installed mobile applications, which are deeply integrated into the
operating system and often receive higher levels of implicit user trust
compared to third-party applications. Ozbay and Bicakci proposed a
security and privacy risk scoring model for pre-installed Android
applications and observed that such applications may operate with
privileged access while receiving less scrutiny from users than
applications downloaded from official app stores~\cite{ozbay2024trust}.

Similarly, Sutter et al.~analyzed firmware images of Android devices and
demonstrated that pre-installed applications may include embedded
third-party components and operate under advantageous system conditions,
which can make their behavior difficult for users to monitor or evaluate
~\cite{sutter2023firmwaredroid}.

This paper proposes a context-aware, user-centric permission governance
framework designed specifically for default mobile applications, where
permission behavior is often less visible to users despite deeper
integration with operating system services. Unlike conventional runtime
permission approaches that maintain access across entire application
sessions, the proposed framework introduces a feature-level
authorization option called ``Allow When Needed,'' which restricts
permission access to the specific functionality that requires it rather
than the duration of application execution.

In addition, the framework incorporates a weighted privacy scoring model
that provides interpretable feedback about cumulative exposure resulting
from repeated permission decisions across interaction scenarios. This
scoring mechanism is intended to support user awareness during
authorization rather than enforce system-level policy restrictions.
Because modifying native mobile operating systems was outside the scope
of this study, the proposed approach was implemented and evaluated via a
web-based simulation platform to enable controlled early-stage
assessment before native mobile deployment thus emulating realistic
permission request scenarios in default applications

The evaluation consists of a cross-sectional survey involving 104
participants to examine user awareness and behavioral patterns regarding
default application permissions, combined with formative usability
testing with eight participants to explore how users interpret and
interact with feature-level authorization and privacy feedback
mechanisms within controlled permission-request scenarios.

The contributions of this work are threefold and focus on exploratory
evaluation of user-centered permission governance mechanisms for default
mobile applications:

\begin{enumerate}
\def\labelenumi{\arabic{enumi}.}
\item
  A feature-based permission governance model tailored for default
  mobile applications that links authorization decisions to individual
  functional interactions rather than application sessions.
\item
  A cumulative privacy scoring mechanism designed to provide
  interpretable feedback about relative exposure associated with
  repeated permission decisions across application scenarios.
\item
  A simulation-based implementation and exploratory empirical evaluation
  examining how feature-level authorization structures influence user
  awareness and decision-making behavior during permission interactions.
\end{enumerate}

The remainder of the paper is organized as follows. Section 2 reviews
related work on mobile permissions and context-aware authorization.
Section 3 presents the proposed framework and simulation design. Section
4 reports the evaluation methodology and results. Section 5 discusses
implications and limitations, and Section 6 concludes the paper.

\section{Literature Review}

\subsection{Mobile Permission Models And Runtime Controls}

Access to sensitive resources like location, camera, microphone, and
storage is controlled with the help of runtime permission frameworks,
which are found in mobile operating systems. According to the Android
Developers documentation, the latest Android operating systems have such
options as one-time access and a possibility to allow only when using
the app, which is directed to minimize the ongoing data exposure
~\cite{androidRuntimePermissions}. These mechanisms represent a shift from earlier install-time
permission models toward more interactive user control.

Although these have been improved, user control remains mostly at the
application or session level. Mendes et al. performed a field study on
user expectations during permission requests in actual field settings
and discovered that a substantial part of the permission access was
unexpected by the user, which revealed that there is a discrepancy
between the intent of the user and the behavior of the system~\cite{mendes2022expectation}.
This issue of persistence comes into play particularly when permissions
that have been granted to one feature are still persisted throughout the
entire application session.

\subsection{User Awareness and Permission Decision Behavior}

Permission governance is highly dependent on user awareness. A study by
Prange et al.~investigated the awareness of the granted permission among
Android users and found that a significant portion of respondents failed
to correctly identify which permissions were granted to their installed
applications~\cite{prange2024feature}. This implies that permission states tend to vanish
when one simply uses the phone in their normal daily use.

The decision-making is also influenced by the quality of permission
explanations. Elbitar et al.~proved that the specifics of permission
rationale words and the way they are presented play a crucial role in
shaping the decisions of the users and their belief in the decisions
made~\cite{elbitar2025power}. Their results support the assertion that permission
systems are not technical controls but also communication, in which
clarity and context determine user behavior.

Wu et al.~also delved into the issue of how user discomfort concerning
permissions is dependent on physical contexts and activities~\cite{wu2025discomfort}.
Their findings show that the perception of privacy risk by users is
situational, and it is possible to argue that the static or
session-based models of permission are not sufficient to describe the
dynamic expectations of users.

\subsection{Fine-Grained and Context-Aware Enforcement Approaches}

Having identified the shortcomings of coarse-grained permission models,
various researchers have considered context-sensitive and fine-grained
enforcement system. Guerra et al.~presented RPCDroid, a runtime analysis
system, that correlates the identifiable dangerous use of permissions
with particular UI events, making it possible to identify various access
contexts within the same application~\cite{guerra2023rpcdroid}. Their contribution draws
our attention to the fact that what is suitable to be permitted in one
feature can be overly permitted in another.

Likewise, Malviya et al.~presented DROIDGEM which categorizes the
permission usage on the per-functional level of a particular app as a
response to user or system events~\cite{malviya2023droidgem}. This strategy facilitates
the interpretation of permission access features on a per-feature basis
and not on an application as a unit.

Most recently, Milanese et al.~tested a permission interpretation model
based on the use of an LLM, which interprets UI context to aid in
providing a user with a decision-making process~\cite{milanese2025llm}. Their article
is explicit that existing mobile platforms are based on permissions on
an application level instead of limiting them to particular features,
allowing them to use more granular governance models.

\subsection{Privacy Concerns In Default And Pre-Installed Applications}

There are unique aspects of privacy, with default and pre-installed
applications, due to their thorough access to the operating system and
their special privileges during deployment. By analyzing firmware images
of large-scale firms using a non-executable, large-scale, and unbiased
method, Sutter et al.~ revealed that pre-installed Android apps may
contain hefty third-party abbreviates and function under advantageous
circumstances~\cite{sutter2023firmwaredroid}. These attributes may make user monitoring
complex.

Ozbay and Bicakci suggested a scoring system to evaluate the privacy and
security risk of pre-installed Android applications, and stated that
pre-installed applications tend to be less scrutinized than third-party
applications downloaded in official stores~\cite{ozbay2024trust}. Innovative analyses
of mobile devices within the emerging markets have also documented
privacy exposure associated to bundle software arrangement and privilege
settings .

Even though the vendors of the platform still enhance the privileged
permission restrictions, such as allowlisting signature permissions of
recent versions of Android, the changes happen mostly at the system
level. They fail to directly discuss how user experience and interpret
permission decisions in their daily interaction with default
applications.

\subsection{Design Implication of Permission Misuse, Ecosystem Evolution, and Design Implications}

Recent work also demonstrates that the permission problem is not a user
interface problem, but an ecosystem evolution problem as well. Wang et
al.~studied bugs related to Android Runtime Permission and discovered
that these bugs are typical among large populations of applications and
may cause unwanted behaviors when permissions are revoked or permission
specifications change across Android versions. This argument motivates
the practical issue that though control mechanisms may be enhanced in
platforms, actual usage may not conform to the intention of the user
because of the actualities of implementation across platforms and OS
versions~\cite{wang2022aper}.

On the ecosystem level, Alkinoon et al. examined the usage trends of
permission over the years and across the app types and found that the
patterns of permission requests change over time and that co-occurrence
of permission sets can uncover significant privacy implications. This
underlines the necessity of governance methods that can be interpreted
by consumers alongside that which withstands the variation of
applications through updates~\cite{alkinoon2025evolving}.

On the developer side developer and end user views of permissions were
compared and, according to Tahaei et al.~developers can request more
substantial permissions because they have not understood permission
scopes, or because the third party libraries require it, whereas users
usually interpret the prompt to mean they are needed. This failure of
connection assists us in understanding why users still encounter
inappropriate demands even in contemporary runtime permission systems,
and it inspires strategies of linking authorization more clearly to
feature particular need instead of the broad access of the app as a
whole~\cite{tahaei2023stuck}.

\subsection{Prompting in Context, Rational Reasoning and Permission Fatigue}

In addition to awareness, the time and manner of permitting influences
quality of decisions. Harbach et al.~conducted research on the large
scale quieting of permission prompts in an on device environment and
demonstrated that the reduction of unnecessary interruptions can enhance
user experience without removing user control. Even though their context
is browser permissions, the mechanism underlying them is applicable to
the context of mobile permission governance as it demonstrates the
conflict between overprompting and significant consent~\cite{harbach2024quieting}.

Elbitar et al.~conducted a detailed study of permission rationales and
assessed the effectiveness of the phrasing of the rationale in
determining the user choice and perceived control. Their results confirm
the practical design requirement that permission prompts should convey
the immediate purpose and context, particularly when users cannot
determine an individual feature to a permission prompt. This is similar
to context anchored authorization, in which explanations are provided in
addition to what the user is executing, rather than what the application
can execute~\cite{elbitar2025power}.

\subsection{Fine Grained Projections between UI Context and Permissions}

An accumulating literature tries to relate permissions to the particular
interaction situation which provokes access to resources. Guerra et
al.~presented the dynamic analysis tool RPCDroid that associates
UI-based events with permissions to access resources sensitive and
proved that applications can use the same sensitive resource with
varying user interactions. This helps in the major assumption that a
single session level grant can be too broad since a user would mean
granting one feature and not others that are using the same session
window~\cite{guerra2023rpcdroid}.

Milanese et al.~examined the fact that large language models can
determine the suitability of permission requests based on UI screen
context, as they essentially view the screen as evidence of an active
feature. Their findings substantiate the practicability of feature
conscious permission judgement, and offer outward validation of forms of
governance scheme that constrain authorization to the active
functionality~\cite{milanese2025llm}.

Independently, Hu et al.~suggested an LLM driven API permission
mapping discovery in the Android framework, which is also motivated by
the observation that incomplete or obsolete permission mappings may
result in wrong permission declaring statements. Though this work is
aimed at finding the right answer, it is applicable to user governance
since erroneous permission declaration renders user facing decisions
meaningless, even with a well-designed UI~\cite{hu2025bamboo}.

\subsection{Software and Firmware Level Transparency Default}

Transparency, where default and pre-installed applications are
concerned, needs more than app store level analysis, it needs firmware
level visibility. Sutter et al.~introduced FirmwareDroid to perform
large scale extraction and static analysis of pre-installed Android
software to restore the transparency of what is shipped on devices and
the behavior of embedded software components. This clarifies the point
that default applications should be given special treatment in terms of
governing since their existence and actions may be more difficult to
monitor or eliminate using normal tools~\cite{sutter2023firmwaredroid}.

Diallo et al.~have explored what they call pre-installed android apps on
low cost devices and have discussed how very cheaply priced phones can
become a distribution channel of privacy violating functionality
particularly in areas where such phones are widespread. This further
inspires permission governance designs which do not presuppose app store
inspection and which work well even when the app is a default unit of
the device environment~\cite{diallo2025budget}.

\subsection{Permission Exposure and Behavioral Drift Empirical Measurement}

The complicated nature of permission exposure in modern mobile
ecosystems is still evidenced by large-scale empirical studies. Ren et
al.~investigated network-based personal information leaking in mobile
traffic and demonstrated that the leakage may happen even when
legitimate applications transmit personal information, but in the form
not intuitive to users~\cite{ren2016recon}. Although the concepts are technically
applied to default applications, more generally this work indicates the
issue of matching the perceived permission grants to the reality of
downstream data flows.

Calciati and Gorla examined updates to Android applications using a
longitudinal study and stated that permission requests tend to widen
with time, especially after major feature releases~\cite{calciati2017evolve}. This
evolutionary trend makes user management more complicated since an
authorization that had been issued with scaled functionality might
subsequently be extended to new behaviors.

Prange et al.~examined users' awareness and control of Android privacy
permissions in a field study and found that revocations often occurred
in bursts and primarily affected rarely used apps or permissions that
were not essential to core functionality~\cite{prange2024feature}. The practice further supports the thesis
that it might be necessary to consider other mechanisms of retrospective
review.

\subsection{Adaptive Privacy Controls and Decision Support Systems}

A number of research papers have been done on adaptive systems to help
users decide on permission. The contextual recommendation mechanisms
presented by Olejnik et al.~ make use of the previous user decisions to
infer authorization preferences in the future~\cite{olejnik2017smarper}. Their findings
show that adaptive systems are able to reduce cognitive load without
destroying user agency.

On the same note, Wijesekera et al.~compared machine learning models that
make permission appropriateness decisions based on contextual metadata
and history of user interactions~\cite{wijesekera2017feasibility}. These systems are also
architecturally distinct from feature-level authorization, but have the
same purpose: eliminating the distance between what is requested and
what the user expects.

Recently, Liu et al.~found that 78.7\% of the recommendations made by
their personalized privacy assistant were adopted and that participants
perceived the assistant and its recommendations as useful and
usable~\cite{liu2016recommendations}. This promotes the need for explainability coupled with
enforcement.

\subsection{Governance in Transparency Mechanisms and Platforms}

The tools of platform level transparency have also developed. Prange et
al.~found that most participants had previously used Android's Permission
Manager, though usually less than once a month, while most Android 12
participants had not used or were unaware of the Privacy Dashboard~\cite{prange2024feature}. This is in concurrence with the
evidence in the past where awareness may not necessarily translate into
long term governance behavior.

Apple App Privacy labels have also been evaluated in the same way.
According to Xiao et al., there were gaps between the disclosed data
practices and the network behavior observed in some types of
applications~\cite{xiao2023lalaine}. Based on these results we can conclude that the
declarative transparency mechanisms are not capable of ensuring
meaningful control.

\subsection{Research Gap and Positioning Of This Study}

Recent research efforts have also made significant progress in enhancing
mobile permission systems with runtime controls, contextual
explanations, and fine-grained analysis of permission usage. Past work
on mobile permission systems has also covered aspects of context-aware
authorization mechanisms, feature-level permission mapping, adaptive
decision support systems, and other aspects that assist users during
permission interaction. Overall, all past work on mobile permission
systems indicates that permission interaction is closely related to user
expectations in terms of context relevance.

However, there are certain limitations associated with past work on
mobile permission systems in terms of user-centric permission
governance. First, much of past work on mobile permission systems is
more focused on system-level or developer-level permission analysis,
permission detection, or permission decision support. Therefore, there
is less emphasis on user-centric permission governance. Second, although
fine-grained permission analysis is part of past work on mobile
permission systems, there is less emphasis on integrating permission
governance with user-centric features in terms of permission
authorization.Third, default or pre-installed mobile applications,
despite their higher privileges and popularity, are an area that is
still unexplored with regard to user-facing permission governance
structures.

The current study seeks to fill in the identified research gaps by
developing a user-centric permission governance model that incorporates
three components within a unified interaction model, namely,
feature-level permission governance through the ``Allow When Needed''
option, contextual permission explanations in association with
functional interactions, and cumulative privacy scores with the
intention of offering interpretable feedback to the user on their
interactions with the permission governance framework. Rather than
developing a new system-level enforcement mechanism, the novelty of the
current study is in developing a unified user-facing interaction model
that is specific to default mobile applications.

Considering the challenges in directly implementing such permission
governance structures within mobile operating systems, the current study
seeks to offer an alternative simulation-based approach to understanding
user interactions with feature-level permission governance structures,
with the intention of offering an understanding of how such structures
can influence user awareness in association with their interactions with
the permission governance framework.

\section{Methods}

\subsection{Research Approach}

The paper utilized a system design and user evaluation
technique to analyze privacy control measures in default mobile
applications. The study entailed creation of a web-based simulation
system deploying a situation-aware permission governance model and
exploratory empirical evaluation with survey data collection and
formative usability testing.

The aim of the methodological design was to determine the effect of the
feature-level authorization and privacy scoring feedback on the user
awareness and decision behavior in the controlled permission request
scenarios.

The overall research process involved parallel data collection and
system development phases. Survey data were collected to examine user
awareness and behavioral patterns, while the simulation platform was
developed to evaluate feature-level authorization within controlled
scenarios. The workflow of the study is illustrated in Fig. 1.

\begin{figure}[t]
  \centering
  \includegraphics[width=\columnwidth]{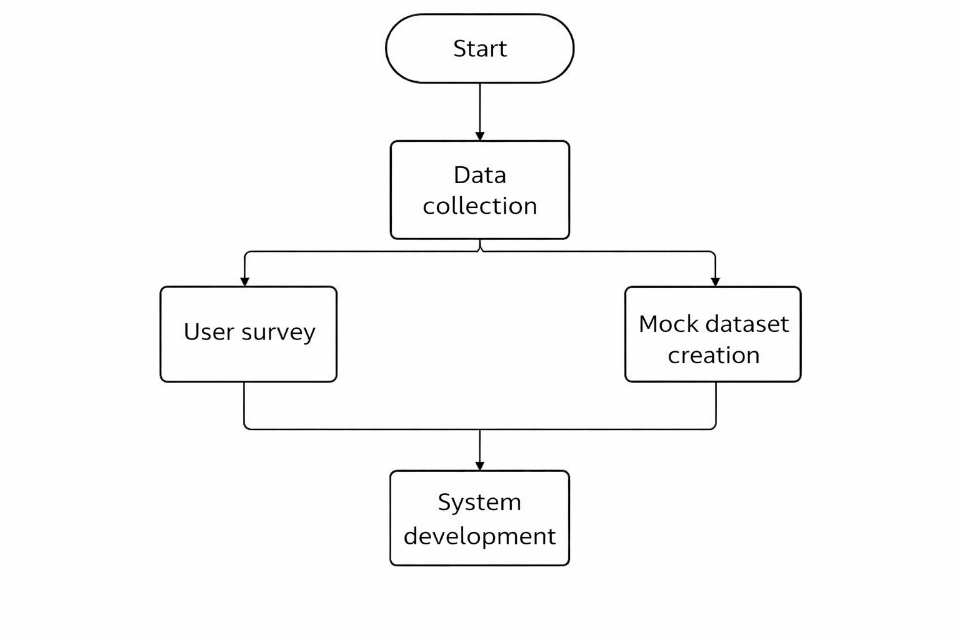}
  \caption{Overview of the Research Design and Data Collection Workflow}
  \label{fig:research-design}
\end{figure}

As shown in Fig. 1, survey analysis informed interpretation of user
behavior patterns, while mock scenario datasets were used to construct
realistic permission interactions within the simulation environment.

\subsection{System Design and Implementation}

The simulation platform is a web-based application
that was created to represent permission interactions that are a common
occurrence in default mobile applications. The system was developed in
such a manner that it imitates realistic permission request prompts and
can compare authorization options in a structured manner.

The system consists of a scenario module responsible for presenting
predefined application tasks, a permission decision interface that
provides four authorization options, a privacy scoring module that
updates cumulative exposure metrics, and a data recording component that
logs user interaction for evaluation purposes.

The internal architecture of the web-based simulation platform is
composed of a backend logic layer, frontend interface layer, encryption
handling module, and integration mechanism. The structural relationship
between components is illustrated in Fig. 2.

\begin{figure}[t]
  \centering
  \includegraphics[width=\columnwidth]{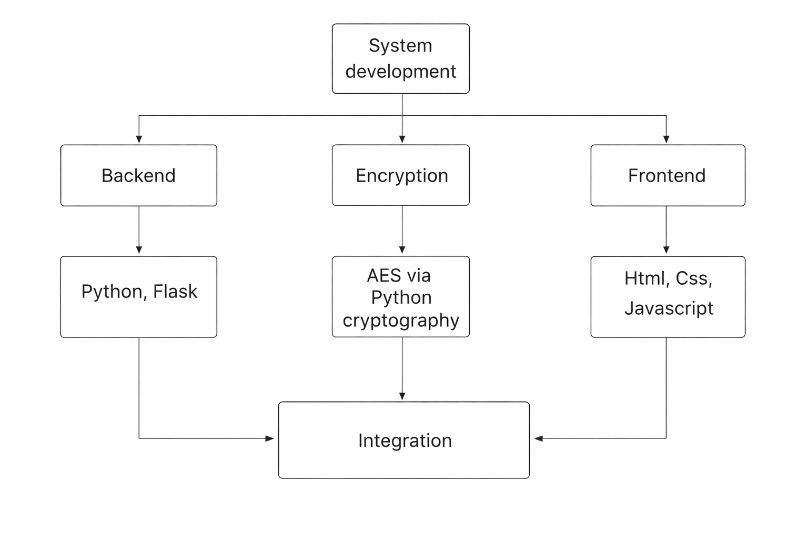}
  \caption{System Architecture and Component Integration}
  \label{fig:system-architecture}
\end{figure}

The backend was implemented using Python and Flask, the frontend
utilized HTML, CSS, and JavaScript, and encryption operations were
handled using AES via the Python cryptography library. These components
were integrated to ensure secure and structured handling of scenario
execution and user interaction logging.

\subsection{Permission Options and Feature-Level Authorization}

The simulation platform offers four permission choices to the
authorization of every permission request:

``Don't Allow,'' ``Allow While Using App,'' ``Allow When Needed,'' and
``Always Allow.''

The ``Allow When Needed'' option represents the feature-level
authorization mechanism proposed in this study. This option is a
conceptual limit in contrast to the session-based authorization, that
preserves permission access during an application session.

\begin{figure}[t]
  \centering
  \includegraphics[width=\columnwidth]{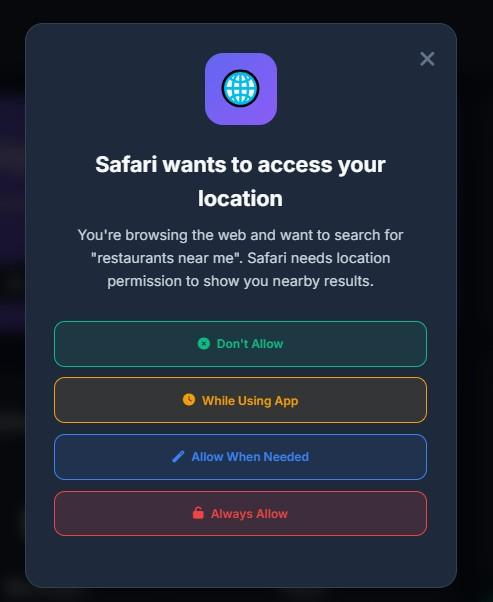}
  \caption{Permission Authorization Options in the Simulation Interface}
  \label{fig:permission-options}
\end{figure}

Figure 3 illustrates the four authorization options presented to users
during each simulated permission request.

\subsection{Privacy Scoring Mechanism}

A direct mathematical model was applied to measure the effects of
permission decisions. A score change was calculated in each permission
interaction according to the sensitivity weight of the permission
requested and the selected authorization modifier.

The score update mechanism is defined as:

\begin{equation}
\text{Score Change} = \text{Permission Weight} \times \text{Action Modifier}
\end{equation}

The cumulative privacy score was successively updated upon each scenario
interaction. The scoring system was intended to give relative feedback
in the simulation environment, and to represent relative exposure to
privacy in various authorization options.

The model is not an absolute privacy risk but works as an assessment
measure to aid user awareness in interaction.

\subsection{Scenario Construction}

To test the framework, structured permission scenarios were created on
the basis of frequent interactions with default mobile applications. Six
default applications were selected for scenario construction, including
Browser, Messages, Maps, Camera, Health, and Phone.

Each application was characterized by 5 task-based scenarios, which made
a total of 30 simulated permission interactions.

Each scenario consisted of a functional task description, the associated
permission request, four authorization options, and a contextual
explanation describing the rationale for access.

The scenario set was to be created to have selected sensitive
permissions such as location access, microphone, camera, contact access,
storage access, and notification permissions.

The constructed scenarios focused on foreground permission requests
triggered by explicit user interactions with application features, while
background or passive permission access behaviors were not included
within the scope of the simulator-based evaluation.

\subsection{Survey Design and Data Collection}

To test the awareness of the user, trust perception and behavioral
pattern about default mobile application permissions, a cross-sectional
survey was carried out. The survey was conducted online with 104 valid
responses.

The questionnaire contained demographic details, habits of device usage,
habits of examining the approval, confidence in default applications,
and the choices of descriptive permission explanations. They had
closed-ended and open-ended questions.

No personal identifiable information was gathered.

\subsection{Participant Demographics}

The demographic characteristics of survey participants were
analyzed to support interpretation of permission awareness and
behavioral response patterns. A total of 104 valid responses were
obtained from users representing multiple age groups and varying levels
of familiarity with smartphone privacy settings.

The majority of respondents (79 participants) were between 18
and 24 years of age, followed by participants aged 25--30 (13
participants), under 18 (5 participants), 35--44 (4 participants), and
45--54 (3 participants). This distribution reflects a participant
population actively engaged in smartphone usage and representative of
common mobile application users.

\begin{figure}[t]
  \centering
  \includegraphics[width=\columnwidth]{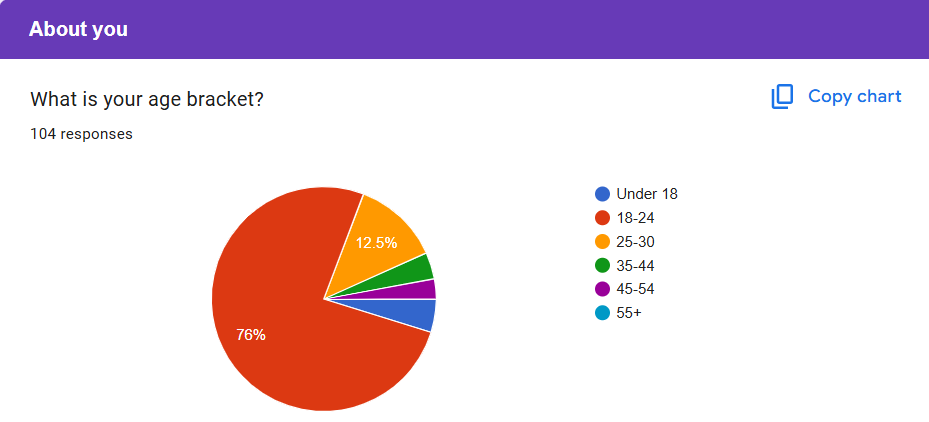}
  \caption{showing the age distribution of participants}
  \label{fig:age-distribution}
\end{figure}

In terms of familiarity with privacy settings, 60 participants
reported being very comfortable managing permission controls, 36
participants indicated basic familiarity, and 8 participants reported
limited familiarity. This variation supported evaluation across users
with different levels of technical confidence.

\begin{figure}[t]
  \centering
  \includegraphics[width=\columnwidth]{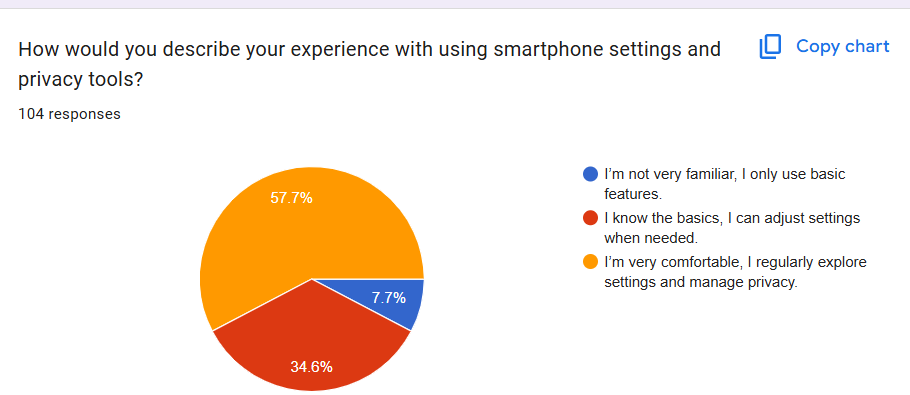}
  \caption{showing the technical expertise of participants}
  \label{fig:technical-expertise}
\end{figure}

Regarding device ecosystem representation, 57 participants
reported using Android devices, 52 reported using iOS devices, and 1
participants reported using both platforms, while 1 participant reported
using a button phone. Inclusion of participants across major mobile
operating systems strengthens the generalizability of the findings
related to permission awareness and interaction behavior.

\begin{figure}[t]
  \centering
  \includegraphics[width=\columnwidth]{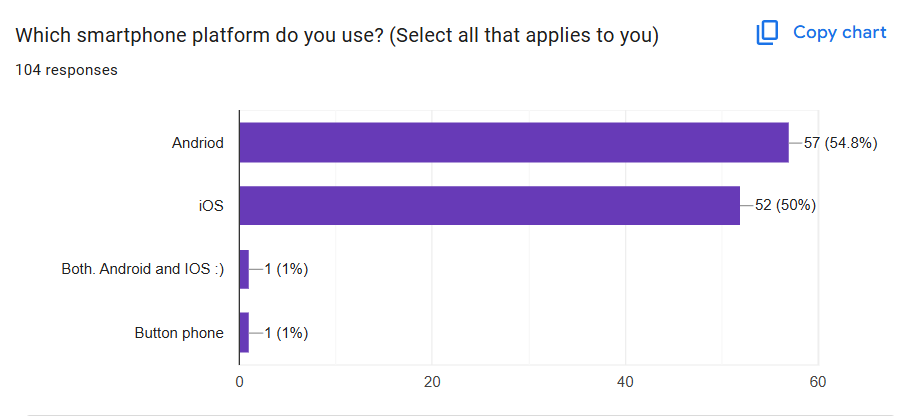}
  \caption{showing the device ecosystem representation}
  \label{fig:device-ecosystem}
\end{figure}

\subsection{Evaluation Procedure}

The evaluation was designed to assess user awareness, behavioral
patterns, and interaction with feature-level permission governance
within a controlled environment. Given the absence of prior user-centric
evaluation frameworks specifically addressing permission governance in
default mobile applications, an exploratory design was adopted as an
appropriate first step. The analysis therefore focuses on identifying
observable trends and usability insights rather than establishing actual
causal relationships or statistically generalizable conclusions.

Survey data were analyzed using descriptive statistical methods such as
frequency distributions and percentage-based summaries to characterize
user awareness, permission review behavior, trust perceptions, and
preferences for contextual explanations. This approach was selected to
provide an initial understanding of user attitudes and behavioral
tendencies related to default mobile application permissions.

In parallel, formative usability testing was conducted with eight
participants to examine how users interpret and interact with the
proposed authorization model and privacy scoring mechanism. The
usability study is intended as a formative evaluation aimed at
identifying interaction patterns, usability challenges, and user
interpretation of feature-level authorization, rather than providing
statistically generalizable results. The sessions focused on qualitative
observation of user behavior, including comprehension of permission
prompts, interpretation of the ``Allow When Needed'' option, and
responses to cumulative privacy feedback..

The use of a web-based simulation platform enabled controlled
presentation of permission scenarios across different application
contexts, ensuring consistency in interaction conditions during
evaluation. However, the findings should be interpreted as preliminary,
as the simulated environment does not fully replicate real-world mobile
usage conditions.

\section{Result}

\subsection{Overview of Respondents}

The following results present descriptive findings from the evaluation
of 104 valid responses. Every respondent filled the main parts of the
questionnaire related to awareness, review behavior, trust perception,
and permission management preferences. There were no unfinished answers
that were not used in statistical reporting.

\subsection{Awareness of Default Application Permissions}

Table 1displays the distribution of answers in terms of awareness of
permissions given to default or pre-installed mobile applications.

\begin{table}[t]
\caption{Awareness of Default Application Permissions (n = 104)}
\label{tab:awareness}
\centering
\begin{tabular}{@{}lrr@{}}
\toprule
Response & Frequency (n) & Percentage (\%) \\
\midrule
Yes & 75 & 72.1 \\
No & 29 & 27.9 \\
\bottomrule
\end{tabular}
\end{table}

Most of the respondents (72.1) stated that they knew the permissions
they gave to default applications on their devices. Nevertheless,
awareness does not necessarily constitute active privacy management
behavior as discussed in the analysis below.

\subsection{Permission Review and Revisit Behavior}

Table 2 presents the frequency of default application permissions review
or modifications by the respondents.

\begin{table}[t]
\caption{Frequency of Permission Review (n = 104)}
\label{tab:review-frequency}
\centering
\begin{tabular}{@{}lrr@{}}
\toprule
Review Frequency & Frequency (n) & Percentage (\%) \\
\midrule
Always (after update) & 4 & 3.8 \\
Often (weekly) & 8 & 7.7 \\
Sometimes (monthly) & 31 & 29.8 \\
Rarely ($\approx$ every 6 months) & 47 & 45.2 \\
Never & 14 & 13.5 \\
\bottomrule
\end{tabular}
\end{table}

Awareness levels were rather high, but there was a lack of consistent
review behavior. Just 11.5\% of the respondents said they would review
permissions frequently or regularly. In contrast, 58.7\% reported either
rarely or never checking default application permissions.

To visually illustrate the distribution of permission review behavior,
Fig. 4 presents the percentage breakdown of respondents' reported review
frequency.

\begin{figure}[t]
  \centering
  \includegraphics[width=\columnwidth]{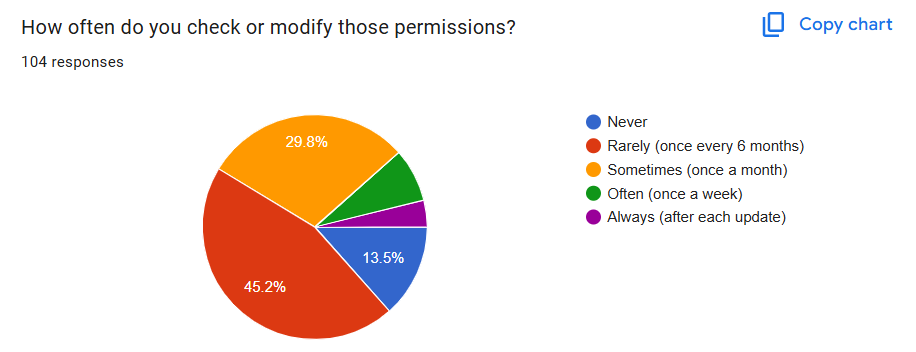}
  \caption{Frequency of Permission Review Among Respondents}
  \label{fig:permission-review}
\end{figure}

The visualization reinforces the concentration of responses in the
``Rarely'' and ``Never'' categories, supporting the observation that
sustained permission monitoring is limited among users.

Table 3 shows revisit behavior before and after application or system
updates.

\begin{table}[t]
\caption{Revisit of Permission Settings After Updates (n = 104)}
\label{tab:revisit-after-updates}
\centering
\begin{tabular}{@{}lrr@{}}
\toprule
Revisit Frequency & Frequency (n) & Percentage (\%) \\
\midrule
Always & 3 & 2.9 \\
Sometimes & 24 & 23.1 \\
Rarely & 40 & 38.5 \\
Never & 37 & 35.6 \\
\bottomrule
\end{tabular}
\end{table}

Over two-thirds of the participants (74.1\%) said they would hardly or
never revisit permission settings once updated. This trend indicates the
lack of behavioral consistency between first-time awareness and
continuous monitoring of permission settings.

\subsection{Trust in Default Applications and Permission Discomfort}

Table 4 is a summary of trust perception and the discomfort experienced
in the pre-request period in the permission requests.

\begin{table*}[t]
\caption{Trust and Discomfort Regarding Default Applications (n = 104)}
\label{tab:trust-discomfort}
\centering
\begin{tabularx}{\textwidth}{@{}Xlrr@{}}
\toprule
Variable & Response Category & Frequency (n) & Percentage (\%) \\
\midrule
Trust Level & Completely & 19 & 18.3 \\
 & A little & 66 & 63.5 \\
 & Not at all & 19 & 18.3 \\
Experienced discomfort about a permission & Yes & 64 & 61.5 \\
 & No & 36 & 34.6 \\
\bottomrule
\end{tabularx}
\end{table*}

The majority respondents have moderate trust, with 63.5\% having
moderate trust in default applications. Simultaneously, 61.5\% indicated
that they have felt uneasy about a permission that had been asked by a
default application. The combination of moderate trust and reported
discomfort point to variability in user perception in the process of the
permission interaction.

\begin{figure}[t]
  \centering
  \includegraphics[width=\columnwidth]{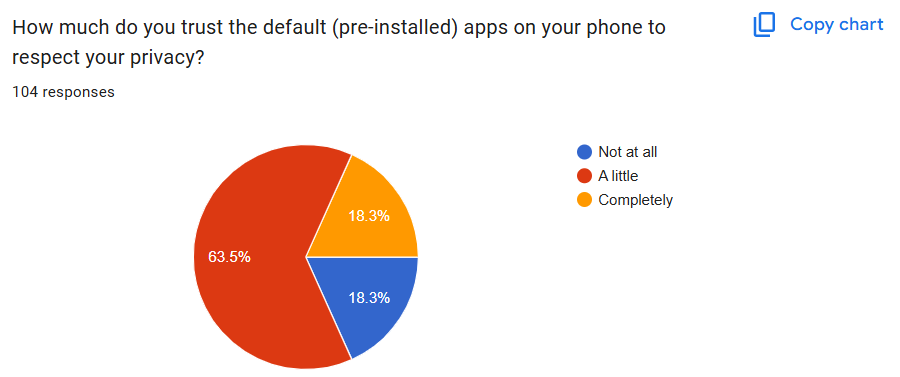}
  \caption{Trust Levels in Default Applications}
  \label{fig:trust-levels}
\end{figure}

Figure 5 provides a graphical representation of reported trust levels in
default mobile applications.

\subsection{Preference for Contextual Explanations and Alerts}

Table 5 summarizes participants' preferences that pertain to contextual
explanations and reminder mechanisms for permission management.

\begin{table*}[t]
\caption{Preference for Contextual Explanations and Permission Alerts}
\label{tab:contextual-explanations}
\centering
\begin{tabularx}{\textwidth}{@{}Xlrr@{}}
\toprule
Variable & Response & Frequency (n) & Percentage (\%) \\
\midrule
Contextual explanation would enhance the comfort & Yes & 60 & 93.8 \\
Reminders or alerts would be desirable & Yes & 81 & 77.9 \\
 & Not sure & 18 & 17.3 \\
 & No & 1 & 1.0 \\
\bottomrule
\end{tabularx}
\end{table*}

Percentage was calculated among those respondents who answered the item.

An overwhelming majority of respondents indicated that contextual
explanations would improve their comfort when granting permissions.
Similarly, most respondents indicated a desire for reminders or alerts
for managing default application permissions effectively. These results
highlight that the users value proactive communication about permissions
and prefer mechanisms that provide clarity before they make any
decisions.

\subsection{Self-Reported Confidence in Permission Management}

The responses indicate a variation in user confidence when managing
permissions for default mobile applications. Although many participants
did report general awareness of permission settings, this did not
consistently translate into confidence during real-time decision-making.

Observations from the usability sessions suggest that uncertainty often
arises at the moment of permission request, particularly when the
purpose and scope of access are not clearly tied to a specific feature.
The participants demonstrated higher confidence when the permission
requests were directly associated with an active task and when feedback
mechanisms provided an immediate interpretability of their choices.

These findings suggest that confidence in permission management is
influenced not only by prior knowledge of settings but also by the
clarity and timing of said permission interactions, supporting the role
of feature-level authorization in reducing uncertainty when decisions
are being made.

\subsection{Consolidated Observations}

Taken together, the results reveal an important gap between user
awareness and sustained privacy management behavior. While a large
proportion of respondents reported being aware of permissions granted to
default applications, most indicated that they rarely revisited
permission settings after updates or during routine device usage. This
pattern suggests that awareness alone does not translate into consistent
privacy-protective behavior.

At the same time, participants expressed moderate trust in default
applications alongside noticeable levels of discomfort when permission
requests were presented without sufficient explanation. The strong
preference for contextual explanations and reminder mechanisms further
indicates that users are more comfortable interacting with permission
systems that provide feature-level transparency rather than persistent
session-level authorization.

These observations collectively support the motivation for the proposed
``Allow When Needed'' framework, which introduces contextual permission
triggers and explanation-aware authorization prompts to improve user
engagement with privacy decisions at the feature level.

\section{Discussion}

The results of this research indicate that there is a consistent pattern
in user behavior regarding permission management in default mobile
applications. Most respondents reported that they were aware of the
permissions granted to pre-installed applications; however, sustained
review and active management of these permissions remained limited. This
suggests that awareness alone does not necessarily translate into
continuous privacy governance behavior. At the same time, a significant
number of respondents reported discomfort with particular permission
requests despite maintaining moderate confidence in default applications
overall. This indicates that user concern is not primarily driven by
general distrust, but rather by uncertainty regarding the scope,
duration, and necessity of permission access in specific situations.

The preference expressed by participants for contextual explanations and
reminder mechanisms further emphasizes the importance of delivering
permission information at the moment of decision-making. These findings
highlight a structural mismatch between existing permission models and
the way users interpret and manage authorization decisions in practice.
The feature-level authorization model proposed in this study intended to
address this mismatch by linking permission activation to individual
functional interactions rather than maintaining access across entire
application sessions. Although the evaluation was conducted within a
simulation environment rather than a native operating system
implementation, the findings provide preliminary evidence that
context-sensitive authorization logic may align more closely with user
expectations and could improve perceived control over privacy-related
decisions.

This research paper has several limitations that should be considered
when interpreting the findings. First, the results rely partially on
self-reported survey responses, which may not always reflect actual user
behavior during real-world permission decision-making. In addition,
although the sample size was appropriate for descriptive behavioral
analysis and exploratory usability evaluation, broader validation with
larger and more diverse participant populations would further strengthen
the generalizability of the findings. Additionally, the complete
reliance on descriptive statistical analysis limits the ability to draw
causal inferences or statistically and reliably validate observed
differences in user behavior.

A further limitation relates to the use of a web-based simulation
environment rather than deployment within a native mobile operating
system. Because participants interacted with permission scenarios in a
controlled experimental setting, their decision-making behavior may have
been influenced by awareness of being observed as part of a study. This
observer effect may lead users to act more cautiously than they would
during routine smartphone use, particularly when responding to
permission requests involving sensitive data access. Future work will
involve implementing the proposed framework within a native mobile
environment and conducting actual controlled experiments using
inferential statistical methods to evaluate its impact on user
decision-making and privacy outcomes under real-world conditions.

In addition, the simulated scenarios focused primarily on foreground
permission requests triggered by explicit user interactions with
application features. However, many default mobile applications perform
background data access operations without direct user awareness,
including passive location tracking, synchronization processes, and
system-level telemetry collection. These forms of invisible permission
usage represent an important dimension of privacy exposure that was not
captured within the scope of the present simulator-based evaluation.
Future extensions of the proposed framework will incorporate monitoring
mechanisms for background permission activity in order to support more
comprehensive protection against passive data access by default
applications.

\section{Conclusion}

This paper has explored the ways in which feature-level authorization
can be used to enhance permission governance of default mobile
applications. An online simulation platform was created that
incorporates an extra authorization policy, ``Allow When Needed'' as
well as a privacy scoring system to give orderly feedback on permission
choices. Optimal framework was tested on 30 simulated permission
condition and research survey data of 104 mobile users.

The findings indicate that the general awareness of default application
permissions exists, but the behavior of active review is not prevalent,
and still, there is a sense of discomfort when a particular permission
is being requested. These results suggest that the current session-based
permission models might not be adequate to meet the user expectations
about the scope and duration of the access. The proposed model offers a
more context-specific alternative to the conventional methods by
limiting permission activation to feature interactions between
individuals.

Even though the evaluation was conducted within a controlled simulation
environment, this study establishes an initial empirical and conceptual
foundation for feature-level permission governance in default mobile
applications. So, rather than presenting conclusive effectiveness, the
findings highlight observable trends in user awareness, their
interaction behavior, and preference for contextual authorization
mechanisms. Future work will focus on implementing the framework within
a native mobile environment and applying rigorous experimental and
statistical methods to validate its impact under real-world conditions.


\bibliographystyle{ACM-Reference-Format}
\bibliography{bib}

@inproceedings{felt2012android,
  author    = {Adrienne Porter Felt and Elizabeth Ha and Serge Egelman and Ariel Haney and Erika Chin and David Wagner},
  title     = {Android Permissions: User Attention, Comprehension, and Behavior},
  booktitle = {Proceedings of the Eighth Symposium on Usable Privacy and Security},
  series    = {SOUPS '12},
  year      = {2012},
  articleno = {3},
  numpages  = {14},
  publisher = {Association for Computing Machinery},
  address   = {New York, NY, USA},
  doi       = {10.1145/2335356.2335360},
  url       = {https://doi.org/10.1145/2335356.2335360}
}

@inproceedings{wijesekera2015remystified,
  author    = {Primal Wijesekera and Arjun Baokar and Ashkan Hosseini and Serge Egelman and David Wagner and Konstantin Beznosov},
  title     = {Android Permissions Remystified: A Field Study on Contextual Integrity},
  booktitle = {24th USENIX Security Symposium (USENIX Security 15)},
  year      = {2015},
  isbn      = {978-1-939133-11-3},
  address   = {Washington, D.C.},
  pages     = {499--514},
  publisher = {USENIX Association},
  month     = aug,
  url       = {https://www.usenix.org/conference/usenixsecurity15/technical-sessions/presentation/wijesekera}
}

@inproceedings{mendes2022expectation,
  author    = {Ricardo Mendes and Andr{\'e} Brand{\~a}o and Jo{\~a}o P. Vilela and Alastair R. Beresford},
  title     = {Effect of User Expectation on Mobile App Privacy: A Field Study},
  booktitle = {2022 IEEE International Conference on Pervasive Computing and Communications (PerCom)},
  year      = {2022},
  pages     = {207--214},
  publisher = {IEEE},
  doi       = {10.1109/PerCom53586.2022.9762379},
  url       = {https://doi.org/10.1109/PerCom53586.2022.9762379}
}

@article{ozbay2024trust,
  author  = {Abdullah {\"O}zbay and Kemal B{\i}{\c c}akc{\i}},
  title   = {Should Users Trust Their Android Devices? A Scoring System for Assessing Security and Privacy Risks of Pre-Installed Applications},
  journal = {ITU Journal of Wireless Communications and Cybersecurity},
  year    = {2024},
  volume  = {1},
  number  = {1},
  pages   = {9--28},
  url     = {https://dergipark.org.tr/en/pub/itujwcc/article/1463270}
}

@inproceedings{sutter2023firmwaredroid,
  author    = {Thomas Sutter and Bernhard Tellenbach},
  title     = {FirmwareDroid: Towards Automated Static Analysis of Pre-Installed Android Apps},
  booktitle = {2023 IEEE/ACM 10th International Conference on Mobile Software Engineering and Systems (MOBILESoft)},
  year      = {2023},
  pages     = {12--22},
  publisher = {IEEE},
  doi       = {10.1109/MOBILESoft59058.2023.00009},
  url       = {https://doi.org/10.1109/MOBILESoft59058.2023.00009}
}

@misc{androidRuntimePermissions,
  author       = {{Android Developers}},
  title        = {Request Runtime Permissions},
  year         = {2026},
  howpublished = {Android Developers documentation},
  url          = {https://developer.android.com/training/permissions/requesting},
  note         = {Accessed 2026-08-28}
}

@inproceedings{prange2024feature,
  author    = {Sarah Prange and Pascal Knierim and Gabriel Knoll and Felix Dietz and Alexander De Luca and Florian Alt},
  title     = {{``I do (not) need that Feature!''}---Understanding Users' Awareness and Control of Privacy Permissions on Android Smartphones},
  booktitle = {Twentieth Symposium on Usable Privacy and Security (SOUPS 2024)},
  year      = {2024},
  isbn      = {978-1-939133-42-7},
  address   = {Philadelphia, PA},
  pages     = {453--472},
  publisher = {USENIX Association},
  month     = aug,
  url       = {https://www.usenix.org/conference/soups2024/presentation/prange}
}

@inproceedings{elbitar2025power,
  author    = {Yusra Elbitar and Alexander Hart and Sven Bugiel},
  title     = {The Power of Words: A Comprehensive Analysis of Rationales and Their Effects on Users' Permission Decisions},
  booktitle = {32nd Annual Network and Distributed System Security Symposium (NDSS 2025)},
  year      = {2025},
  publisher = {Internet Society},
  doi       = {10.14722/ndss.2025.230544},
  url       = {https://www.ndss-symposium.org/ndss-paper/the-power-of-words-a-comprehensive-analysis-of-rationales-and-their-effects-on-users-permission-decisions/}
}

@inproceedings{wu2025discomfort,
  author    = {Yuxi Wu and Jacob Logas and Devansh Ponda and Julia Haines and Jiaming Li and Jeffrey Nichols and W. Keith Edwards and Sauvik Das},
  title     = {Modeling End-User Affective Discomfort With Mobile App Permissions Across Physical Contexts},
  booktitle = {Symposium on Usable Security and Privacy (USEC 2025)},
  year      = {2025},
  publisher = {Internet Society},
  url       = {https://www.ndss-symposium.org/ndss-paper/auto-draft-592/}
}

@inproceedings{guerra2023rpcdroid,
  author    = {Michele Guerra and Roberto Milanese and Rocco Oliveto and Fausto Fasano},
  title     = {RPCDroid: Runtime Identification of Permission Usage Contexts in Android Applications},
  booktitle = {Proceedings of the 9th International Conference on Information Systems Security and Privacy (ICISSP 2023)},
  year      = {2023},
  pages     = {714--721},
  publisher = {SciTePress},
  isbn      = {978-989-758-624-8},
  doi       = {10.5220/0011797200003405},
  url       = {https://doi.org/10.5220/0011797200003405}
}

@inproceedings{malviya2023droidgem,
  author    = {Vikas K. Malviya and Yan Naing Tun and Chee Wei Leow and Ailys Tee Xynyn and Lwin Khin Shar and Lingxiao Jiang},
  title     = {Fine-Grained In-Context Permission Classification for Android Apps Using Control-Flow Graph Embedding},
  booktitle = {2023 38th IEEE/ACM International Conference on Automated Software Engineering (ASE)},
  year      = {2023},
  pages     = {1225--1237},
  publisher = {IEEE},
  doi       = {10.1109/ASE56229.2023.00056},
  url       = {https://doi.org/10.1109/ASE56229.2023.00056}
}

@inproceedings{milanese2025llm,
  author    = {Roberto Milanese and Michele Guerra and Michele Daniele and Giovanni Fabbrocino and Fausto Fasano},
  title     = {Assessing the Effectiveness of an LLM-Based Permission Model for Android},
  booktitle = {Proceedings of the 11th International Conference on Information Systems Security and Privacy (ICISSP 2025)},
  year      = {2025},
  volume    = {2},
  pages     = {36--47},
  publisher = {SciTePress},
  doi       = {10.5220/0013128100003899},
  url       = {https://doi.org/10.5220/0013128100003899}
}

@inproceedings{wang2022aper,
  author    = {Sinan Wang and Yibo Wang and Xian Zhan and Ying Wang and Yepang Liu and Xiapu Luo and Shing-Chi Cheung},
  title     = {APER: Evolution-Aware Runtime Permission Misuse Detection for Android Apps},
  booktitle = {Proceedings of the 44th International Conference on Software Engineering},
  series    = {ICSE '22},
  year      = {2022},
  pages     = {125--137},
  publisher = {Association for Computing Machinery},
  address   = {New York, NY, USA},
  doi       = {10.1145/3510003.3510074},
  url       = {https://doi.org/10.1145/3510003.3510074}
}

@article{alkinoon2025evolving,
  author  = {Ali Alkinoon and Trung Cuong Dang and Ahod Alghuried and Abdulaziz Alghamdi and Soohyeon Choi and Manar Mohaisen and An Wang and Saeed Salem and David Mohaisen},
  title   = {A Comprehensive Analysis of Evolving Permission Usage in Android Apps: Trends, Threats, and Ecosystem Insights},
  journal = {Journal of Cybersecurity and Privacy},
  year    = {2025},
  volume  = {5},
  number  = {3},
  articleno = {58},
  doi     = {10.3390/jcp5030058},
  url     = {https://doi.org/10.3390/jcp5030058}
}

@inproceedings{tahaei2023stuck,
  author    = {Mohammad Tahaei and Ruba Abu-Salma and Awais Rashid},
  title     = {Stuck in the Permissions With You: Developer \& End-User Perspectives on App Permissions \& Their Privacy Ramifications},
  booktitle = {Proceedings of the 2023 CHI Conference on Human Factors in Computing Systems},
  series    = {CHI '23},
  year      = {2023},
  articleno = {168},
  publisher = {Association for Computing Machinery},
  address   = {New York, NY, USA},
  doi       = {10.1145/3544548.3581060},
  url       = {https://doi.org/10.1145/3544548.3581060}
}

@inproceedings{harbach2024quieting,
  author    = {Marian Harbach and Igor Bilogrevic and Enrico Bacis and Serena Chen and Ravjit Uppal and Andy Paicu and Elias Klim and Meggyn Watkins and Balazs Engedy},
  title     = {Don't Interrupt Me---A Large-Scale Study of On-Device Permission Prompt Quieting in Chrome},
  booktitle = {31st Annual Network and Distributed System Security Symposium (NDSS 2024)},
  year      = {2024},
  publisher = {Internet Society},
  url       = {https://www.ndss-symposium.org/ndss-paper/dont-interrupt-me-a-large-scale-study-of-on-device-permission-prompt-quieting-in-chrome/}
}

@misc{hu2025bamboo,
  author        = {Han Hu and Wei Minn and Yonghui Liu and Jiakun Liu and Ferdian Thung and Terry Yue Zhuo and Lwin Khin Shar and Debin Gao and David Lo},
  title         = {Bamboo: LLM-Driven Discovery of API-Permission Mappings in the Android Framework},
  year          = {2025},
  eprint        = {2510.04078},
  archiveprefix = {arXiv},
  primaryclass  = {cs.SE},
  doi           = {10.48550/arXiv.2510.04078},
  url           = {https://arxiv.org/abs/2510.04078}
}

@misc{diallo2025budget,
  author        = {Alioune Diallo and Anta Diop and Abdoul Kader Kabore and Jordan Samhi and Aleksandr Pilgun and Tegawend{\'e} F. Bissyande and Jacque Klein},
  title         = {Security Evaluation of Android Apps in Budget African Mobile Devices},
  year          = {2025},
  eprint        = {2509.18800},
  archiveprefix = {arXiv},
  primaryclass  = {cs.CR},
  doi           = {10.48550/arXiv.2509.18800},
  url           = {https://arxiv.org/abs/2509.18800}
}

@inproceedings{ren2016recon,
  author    = {Jingjing Ren and Ashwin Rao and Martina Lindorfer and Arnaud Legout and David Choffnes},
  title     = {ReCon: Revealing and Controlling PII Leaks in Mobile Network Traffic},
  booktitle = {Proceedings of the 14th Annual International Conference on Mobile Systems, Applications, and Services},
  series    = {MobiSys '16},
  year      = {2016},
  pages     = {361--374},
  publisher = {Association for Computing Machinery},
  address   = {New York, NY, USA},
  doi       = {10.1145/2906388.2906392},
  url       = {https://doi.org/10.1145/2906388.2906392}
}

@inproceedings{calciati2017evolve,
  author    = {Paolo Calciati and Alessandra Gorla},
  title     = {How Do Apps Evolve in Their Permission Requests? A Preliminary Study},
  booktitle = {2017 IEEE/ACM 14th International Conference on Mining Software Repositories (MSR)},
  year      = {2017},
  pages     = {37--41},
  publisher = {IEEE},
  doi       = {10.1109/MSR.2017.64},
  url       = {https://doi.org/10.1109/MSR.2017.64}
}

@inproceedings{olejnik2017smarper,
  author    = {Katarzyna Olejnik and Italo Dacosta and Joana Soares Machado and K{\'e}vin Huguenin and Mohammad Emtiyaz Khan and Jean-Pierre Hubaux},
  title     = {SmarPer: Context-Aware and Automatic Runtime-Permissions for Mobile Devices},
  booktitle = {2017 IEEE Symposium on Security and Privacy (SP)},
  year      = {2017},
  pages     = {1058--1076},
  publisher = {IEEE},
  doi       = {10.1109/SP.2017.25},
  url       = {https://doi.org/10.1109/SP.2017.25}
}

@inproceedings{wijesekera2017feasibility,
  author    = {Primal Wijesekera and Arjun Baokar and Lynn Tsai and Joel Reardon and Serge Egelman and David Wagner and Konstantin Beznosov},
  title     = {The Feasibility of Dynamically Granted Permissions: Aligning Mobile Privacy with User Preferences},
  booktitle = {2017 IEEE Symposium on Security and Privacy (SP)},
  year      = {2017},
  pages     = {1077--1093},
  publisher = {IEEE},
  doi       = {10.1109/SP.2017.51},
  url       = {https://doi.org/10.1109/SP.2017.51}
}

@inproceedings{liu2016recommendations,
  author    = {Bin Liu and Mads Schaarup Andersen and Florian Schaub and Hazim Almuhimedi and Shikun (Aerin) Zhang and Norman Sadeh and Yuvraj Agarwal and Alessandro Acquisti},
  title     = {Follow My Recommendations: A Personalized Privacy Assistant for Mobile App Permissions},
  booktitle = {Twelfth Symposium on Usable Privacy and Security (SOUPS 2016)},
  year      = {2016},
  isbn      = {978-1-931971-31-7},
  address   = {Denver, CO},
  pages     = {27--41},
  publisher = {USENIX Association},
  month     = jun,
  url       = {https://www.usenix.org/conference/soups2016/technical-sessions/presentation/liu}
}

@inproceedings{xiao2023lalaine,
  author    = {Yue Xiao and Zhengyi Li and Yue Qin and Xiaolong Bai and Jiale Guan and Xiaojing Liao and Luyi Xing},
  title     = {Lalaine: Measuring and Characterizing Non-Compliance of Apple Privacy Labels},
  booktitle = {32nd USENIX Security Symposium (USENIX Security 23)},
  year      = {2023},
  isbn      = {978-1-939133-37-3},
  address   = {Anaheim, CA},
  pages     = {1091--1108},
  publisher = {USENIX Association},
  month     = aug,
  url       = {https://www.usenix.org/conference/usenixsecurity23/presentation/xiao-yue}
}

\appendix 

\section{Open Science} 

\textbf{\href{https://www.kaggle.com/datasets/subhajournal/trojan-detection}{Availability of Data and Materials:}} The survey dataset generated and analyzed during the current study is available from the corresponding author upon reasonable request

\section{Ethical Considerations}

\textbf{Ethics Approval:} This study involved voluntary survey participation. No personally identifiable information was collected.

\textbf{Conflicts of Interest:} The authors declare no conflicts of interest to report regarding the present study.

\end{document}